# ANOMALOUS CONTRAST AS AN ADAPTIVE VIOLATION OF THE TALBOT-PLATEAU LAW

Ernest Greene and Jack Morrison
Department of Psychology, University of Southern California
Los Angeles, California, USA

Corresponding author:  Ernest Greene   egreene@usc.edu
Running title:  Anomalous contrast as Talbot-Plateau violation

ABSTRACT

Purpose:  To better understand anomalous contrast mechanisms that allow flicker-fused stimuli to be visible even when they provide the same average luminance as background.

Method: Stimulus flicker was used to elicit differential activation of ON and OFF retinal channels at frequencies above the flicker-fusion threshold.  Providing balanced light energy to ON and OFF channels will normally cause the stimulus to vanish into the background.

Results:  We used ultra-brief bright pulses, combined with ultra-long dark pulses, to elicit "anomalous contrast" that rendered the stimulus visible, even though it had the same average luminance as the background.  The duration and intensity of flicker components were varied to gain insight into the conditions that would elicit this effect.

Conclusions:  Anomalous contrast displays violated the Talbot-Plateau law, but in doing so, provided an adaptive way to register and signal contours that matched background luminance. These findings contribute additional details about this visual adaptation, and we discuss how the retinal circuitry provides for stimulus visibility.

## 1. INTRODUCTION

*[T}he striking fact that a deflection in the same direction takes place on illumination and on darkening, suggests that there are in the eye two or more different processes occurring partly simultaneously, partly successively, whose fusion determines the form of the electrical reaction.*
*Einthoven & Jolly, 1909* [1]

We have long thought of visibility of a stimulus as being determined by the amount of light it delivers.  It appears bright if it emits more light than the scene that surrounds it, and dark if less is provided.  One's theory of perception can be built on a unidimensional model wherein brightness and discrimination judgments are based on the relative abundance of light.  The eye would register and signal a bright stimulus by increasing the level of optic-nerve activity, and signal the dark stimulus with a decrease in activity level.

The turning point for a different conceptual model was provided early in the 20th century by investigators who noted differential activation of retinal neurons by increments and decrements of light.[1-5]  These became known as "ON" and "OFF" channels, with bipolar and ganglion cells being central players in the dual-channel concept.[6]   Rather than reducing the level of optic nerve activation in response to a dark stimulus, there was selective activation of specific nerve fibers that could elicit a conscious perception that it was darker than the surrounding scene.

Now, more than a century of perceptual and physiological work has firmly documented the role of ON and OFF retinal channels in registering differentials in light level.  Even so, the mechanisms are not fully understood.  Recent work from our laboratory points to channel interactions that were not expected.[7,8]   We displayed flicker-fused letters as dot patterns on an LED array.  Background dots surrounding a letter pattern had steady emission, and flicker frequency of the letter pattern was above the fusion threshold, so they also appeared steady.  We used the Talbot-Plateau law [9] to determine the average luminance of the flicker.  When the background and letter both manifested the same average luminance, they should appear equally bright, and the letter should be invisible.

However, the letters could be perceived and identified when the flicker was produced using ultrabrief flashes.[7,8]  The ultrabrief flashes induced what we have described as "anomalous contrast," this being a perceived differential in stimulus luminance even the average luminance of the stimulus matches the background.  This is a violation of the Talbot-Plateau law, to our knowledge, the only violation that cannot be attributed to invalid specification of flicker measures .  We have speculated that the mismatch of duration and/or intensity of dark and bright flicker components can modify ON- and OFF-channel interactions, promoting competition between the channels rather than combined activity that yields perception of the luminance average predicted by the Talbot-Plateau law.

The present work produced flicker with bright and dark departures from background, designated as pulses.  We first replicated the basic conditions that had produced anomalous contrast, showing that the effects can be seen with a larger font size and where the letter is outlined by a single-file array of dots. Then several tasks manipulated frequency, number of cycles, amplitude of pulses, and duration of pulses to better determine the relative contribution of these factors to the anomalous contrast violation. We discuss how the ON and OFF channels normally determine average luminance of a flickering stimulus, but will register and signal anomalous contrast when there is an extreme imbalance of dark/bright duration and/or intensity. In closing, we outline how the anomalous contrast violation might contribute to registering and signaling image features, and therefore can be considered to be a beneficial adaptation.

2. METHODS

*2.1 Display Equipment*

Stimuli were displayed on a 64x64 array of red LEDs that was custom designed and fabricated by Digital Insight. These LEDs have peak emission at 633 nm. A given LED provides 90% of its light within a span of 5.3 mm, which is designated as the dot diameter. Center-to-center spacing was 4.83 mm and the total span of the array was 309 mm. The array was placed at a viewing distance of 1.5 m, so the visual angles for dot diameter, spacing, and span were 12, 11, and 707 minutes, respectively. Rise/fall time of emission of these LEDs is 3 ns. Timing precision for display of stimuli was 1μs. A compact Windows PC, custom programmed with Tcl/Tk applications, provided experimental control of stimulus displays and recording of respondent judgments.

*2.2 Letter Design and Display*

Letters were displayed as dot patterns, styled as Arial True-Type 60 point fonts, either with solid fill or as unfilled outlines. (Examples are provided in the Results section, in combination with models from the first task.) Letters were 32 dots tall. Solid letters had a stroke width of 5 dots, and single-file dots provided the boundaries of outline letters. Depending on contrast level and other treatment conditions, letters could appear bright or dark against a uniform background. All letter displays were against a steady background of luminance at 8 Cd/m$^2$ that extended across the full array.

Most of the tasks used displays with bright-pulse durations of 2000 μs and/or 2 μs. For 250 Hz displays, this corresponds to 50% and 0.05% duty cycles, respectively. To provide letters with average luminance that matched the background, i.e., balanced luminance, bright-pulse intensity for the 2000 μs displays was 8 Cd/m$^2$, matching the 2000 μs of zero light emission for the dark pulse. For the 2 μs flicker, dark-pulse duration was 3998 μs (maintaining the full flicker period of 4000 μs). Bright-pulse intensity was adjusted to compensate for the absence of light during the extended duration of the dark pulse.

Tasks 1 and 7 displayed solid-fill and outline letters, and all other tasks displayed only outline letters. All letters in Tasks 2-6 were luminance-balanced, and all tasks except Task 2 displayed the letters at 250 Hz.

*2.3 Task 1: Anomalous Contrast Replication*

This task displayed solid-fill and outline Arial 60 letters. Each letter was displayed at 250 Hz for 300 ms, with bright-pulse durations of 2 μs or 2000 μs. Bright-pulse intensity was varied across a range from 75% to 125%, which brackets the intensity that provides luminance balance (100%) Bright-pulse intensities were chosen at random, weighted for heavier sampling in the 100% vicinity. The session for a given respondent provided 400 displays, half with 2 μs flicker and half at 2000 μs flicker.

*2.4 Task 2: Perceived flicker, visibility, and recognition as a function of flicker frequency*

This task displayed luminance-balanced letters for 300 ms, using either 50% or 0.05% duty cycle. Flicker frequency was varied from 25 to 250 Hz in 25 Hz increments. The ten levels of frequency and two levels of duty cycle provided 20 treatment combinations. Each combination was presented for 20 trials, for a total of 400 display trials.

*2.5 Task 3: Letter recognition as a function of number of cycles displayed*

For this task, the duration of the display on a given trail was determined by the number of flicker cycles, rather than being displayed for 300 ms. Letters were displayed with bright-pulse durations of 2000 and 2 μs, varying the number of cycles provided on a given trial, specifically: 1, 2, 4, 8, 16, 32, 64, 128. The longest display (128 cycles) was 512 ms in duration. The 16 treatment combinations (eight levels of cycle number and two levels of bright-pulse duration) were each displayed for 25 trials for a total of 400 trials for a given session.

*2.6 Task 4: Letter recognition as a function of percent departure from background*

This task displayed all letters with 2 μs balanced-flicker. Amplitude of bright- and dark-pulse departure from background intensity was specified as a percentage of the dark-pulse range. Total non-emission was designated as 100%, and the protocol provided 10 levels from 10% to 100%, i.e., 10, 20, 30 ..... 90, 100. All displays were for 32 cycles (128 ms). The ten dark-pulse levels were each displayed for 40 trials, for a total of 400 session trials.

### 2.7 Tasks 5 and 6: Letter recognition as a function of dark-pulse duration

These tasks also displayed all letters with 2 μs balanced-flicker. For each display there was 100% dark departure from background, i.e., zero emission, but the duration of the dark pulse was varied. The 4000 μs duration of the flicker period was considered to be 100%, and the durations of dark pulse were varied in 10% increments, i.e., 10, 20, 30,.....90, 100. For Task 5 the dark pulse followed immediately on offset of the bright pulse. For Task 6 the dark pulse preceded the bright pulse. All displays were for 32 cycles (128 ms). Each of the ten levels of dark pulse was displayed for 40 trials, for a total of 400 trials for the session.

### 2.8 Task 7: Eliciting Anomalous Contrast Recognition with Movement

To test the hypothesis that movement of a luminance-balanced letter would elicit anomalous contrast recognition, this task followed the basic design of Task 1, but replaced 2 μs flicker with a motion-inducing treatment. Outline Arial 60 letters were displayed at 100% (luminance balanced) bright-pulse intensity across eight levels of cycle, i.e., flicker repetitions at 1, 2, 4.... 128 cycles. All letters were displayed with a bright-pulse duration of 2000 μs (50% duty cycle). Half of the letters were displayed at the fixed central location, these being designated as the "still" letters. The other half were displayed with motion, wherein the full pattern of the letter moved one position in a randomly selected cardinal or diagonal direction every 10,000 μs. The pattern would move for three steps (dots) in the same direction before reversing for three steps to return to the starting location. Perceptually, this was perceived as a "jiggle" of the letter, and that term will be used to describe the motion. The two motion conditions (static and jiggle) were displayed at the 8 levels of cycle number for 25 trials per combination, for a total of 400 trials.

*2.9 Experimental Approval and Execution*

Experimental protocols were approved by the University of Southern California, University Park Institutional Review Board (FWA 000007099).  Fifty two respondents were recruited from the Psychology Subject Pool, or by posted flyer, the latter being paid for their participation.  Ages ranged from 18-32, and by self-report there were 21 males and 31 females.  Informed consent was obtained for all the participants involved in the study.  Each was informed of the task requirements, and that they could discontinue participation at any time without penalty.  Each was tested for visual acuity, but in no instance did this appear to be a factor in task performance.

Each respondent was tested individually on one and only one task.  Twelve respondents provided the data for Task 1, and ten were tested for Tasks 5 and 7.  Five respondents provided the data for each of the other tasks.

The test room was illuminated by a single 7 Watt DC bulb (6000K), with pulse-width modulation that provided 10 lux of steady ambient light.  The respondent was seated against a wall, with the bulb positioned directly above, at 2.25 m from the floor.  These lighting conditions had previously been found to produce an average pupil diameter of 6.66 mm.

The LED display was positioned in the line of sight of the respondent at a distance of 1.5 m.  Respondents viewed the display with both eyes open.  Background luminance was maintained throughout the session.  On each trial a letter was randomly chosen from the alphabet, and a brief tone was provided at onset of the display in case the letter was imperceptible.  All tasks asked for letter recognition, and respondents were expected to name the letter or say that no letter had been perceived.  For Task 2 they were also asked to say if the letter appeared to flicker.  The experimenter entered the response on the computer, and then launched the next trial.  Neither the respondent nor the experimenter was provided with feedback about whether letters were correctly identified.  Most respondents completed the session in about 45 minutes.

*2.10 Statistical Analysis*

As required by the specific task, correct identification of letters and reporting that flicker was perceived were scored as 1; incorrect letter identification, lack of flicker, and invisibility (no letter perceived) were scored as 0. Logistic regression models were constructed to reflect probabilities of letter recognition, visibility, and/or flicker perception.

All seven tasks used mixed-effects repeated-measures logistic regression models, treating respondents as random effects.  All models provided 95% confidence intervals for the estimated probabilities at each treatment level, which provide the robust measure of uncertainty.

For Tasks 1 and 7, modeling was done at each bright-pulse intensity using B-Spline forms with interactions included as fixed effects.]10,11]  This provided flexibility to capture the U-shape of the data. Nonparametric bootstrap methods with 1,000 samples were used to generate 95% confidence intervals for the estimated probabilities.

For flicker-fusion judgments in Task 2, all responses that no letter was seen were excluded before the probability calculation and data modeling.  For Task 4 and 6, log transformation was applied to the luminance levels (10 unique values from 10% to 100%) to make the model more stable and interpretable. Once these models were derived, abscissa values were returned to linear scale when creating the plot to better reflect the rapid rise in probability of recognition that was evident in the raw data.

## 3. RESULTS

### 3.1 Terms of Art

The displays provided flicker as a departures from steady background luminance, comprised of "bright pulses" and "dark pulses."  We are avoiding the term "flash," for the world may not be ready for the concept of "dark flashes."  We may use the pulse duration for labeling a given display, e.g., a "2 μs flicker."  In this context, the "flicker" label specifies the treatment conditions and does not imply that the display was visible or that observers perceived stimulus pulsations.  The terms "flicker period" or "flicker cycle" refers to the duration of the pulse pair.  The term "duty cycle" specifies the duration of the bright pulse as a percentage of flicker period.  A dark pulse is assumed to complete the period unless otherwise specified.  The term "luminance balanced" is used to describe displays wherein the duration and intensity of dark and bright pulses provided an average luminance equal to the steady background luminance.  All letters were luminance balanced on Tasks 2-7, but many pulse combinations were deliberately unbalanced on Task 1.

*3.2 Anomalous Contrast Replication (Task 1)*

This task was conducted to confirm that solid-fill and outline Arial 60 letters (illustrated in Fig 1A) would show the same anomalous luminance contrast effect that was found with Arial 33 letters.[7,8]  The anomalous contrast effect can be demonstrated with a flicker frequency of 250 Hz, bright-pulse durations of 2000 μs and 2 μs (50% and 0.05% duty cycle), and testing with a range of pulse intensities that bracket the Talbot-Plateau prediction of balanced luminance.[7,8] On each side of the predicted balance point, designated as 100%, the pulse combinations were progressively weighted in favor of the dark or bright pulse.  This yielded various degrees of perceptibility as reflected in the probability of recognition.  In the vicinity of the predicted balance point, i.e., 100% contrast, the letters have an average luminance that matches background luminance, so in theory they should not be visible.  The earlier work found that with the frequency at 250 Hz, recognition of letters at 100% contrast was only slightly above chance level (4%) with bright-pulse duration at 2000 μs.  But for displays that used bright-pulse durations of 2 μs, most of the letters were identified.

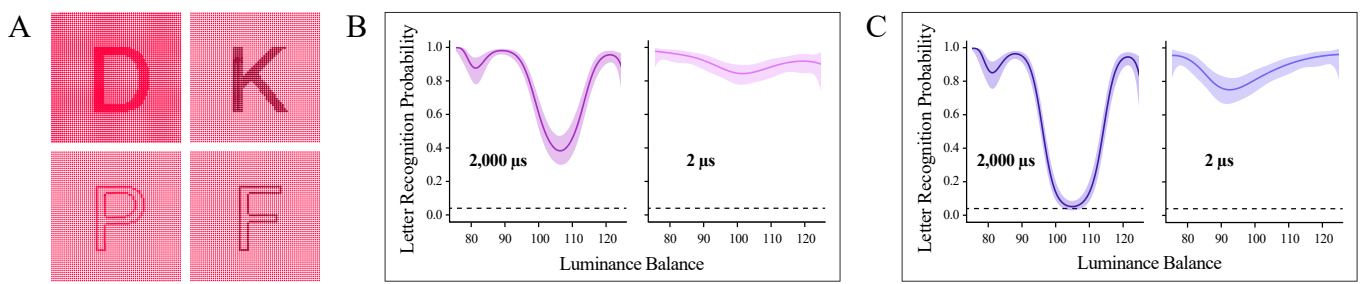

Figure 1.  A. Solid-fill and outline Arial 60 letters were used in a contrast discrimination task that was designed to produce anomalous contrast.  B. Solid-fill letters that were displayed with 2000 μs flicker were readily identified at each end of the contrast range, but many of the letters were not visible where the average luminance of letters was at or close to background luminance. Letter displays that used 2 μs flicker were identified with high probability across the full range, including where the letters and background were luminance balanced.  C. For outline letters that were displayed with 2000 μs flicker, recognition for luminance-balanced trials was near chance. The letters that were displayed with 2 μs flicker were identified with high probability even when they were luminance balanced, which we attribute to activation of anomalous contrast mechanisms.

This replication protocol displayed the solid-fill and outline letters at 250 Hz with 2000 μs and 2 μs flicker.  The panels of Figure 1B and 1C show the logistic regression models for probability of recognition.  Recognition for outline letters that were displayed with 2000 μs flicker was in the chance range when average luminance was close to background luminance, i.e., 100% contrast. This is consistent with what was previously found with the smaller (and solid-fill) Arial 33 letters. The solid-fill letters also manifested depressed recognition in the vicinity of the balance point, but about 40% of the letters were identified. The difference in depth of the non-recognition depression between solid-fill and outline letters is likely due to the sparse nature of the outline pattern. This issue might be more fully examined at a later time.

As expected, the 2 μs flicker condition substantially elevated recognition of the letters in the vicinity of balanced luminance. At the balance point predicted by Talbot-Plateau, i.e., 100% contrast, minimum recognition for outline letters was roughly 75% and the solid-fill letters were identified on about 80% of the trials. This essentially replicates the earlier finding that ultrabrief bright pulses can render flicker-fused letters as visible when their average luminance matches the luminance of the background within which they lie.

*3.3 Perceived Flicker, Recognition, and Visibility as a Function of Frequency (Task 2)*

Task 1 varied average contrast of letters, providing physical differences in contrast that ranged from dark to bright, with the middle (100%) condition manifesting luminance balance. Beginning with Task 2, all remaining tasks displayed only luminance-balanced letters on every trial, irrespective of whatever other treatment condition was being manipulated. Also, except for the final task (Task 7), only outline letters were displayed.

For Task 2, frequency of flicker-fused letter displays was varied from 25 Hz to 250 Hz, each display using a 50% or 0.05% duty cycle. Respondents were asked to report if the letter appeared as steady emission, or manifested a flicker. They also reported if the letter was absent on a given trial, i.e., not visible, and they named any letter that was seen.

Figure 2A provides logistic group models that show probabilities of flicker fusion responses, letter recognition, and visibility. For letters displayed at 50% duty cycle, one can see a monotonic decline in flicker-fusion judgments. Flicker was reliably perceived on almost every trial at 25 Hz, but dropped to nearly zero perceived flicker at 250 Hz. Probability that the letter was visible had nearly the same profile of decline, as did the model for recognition.

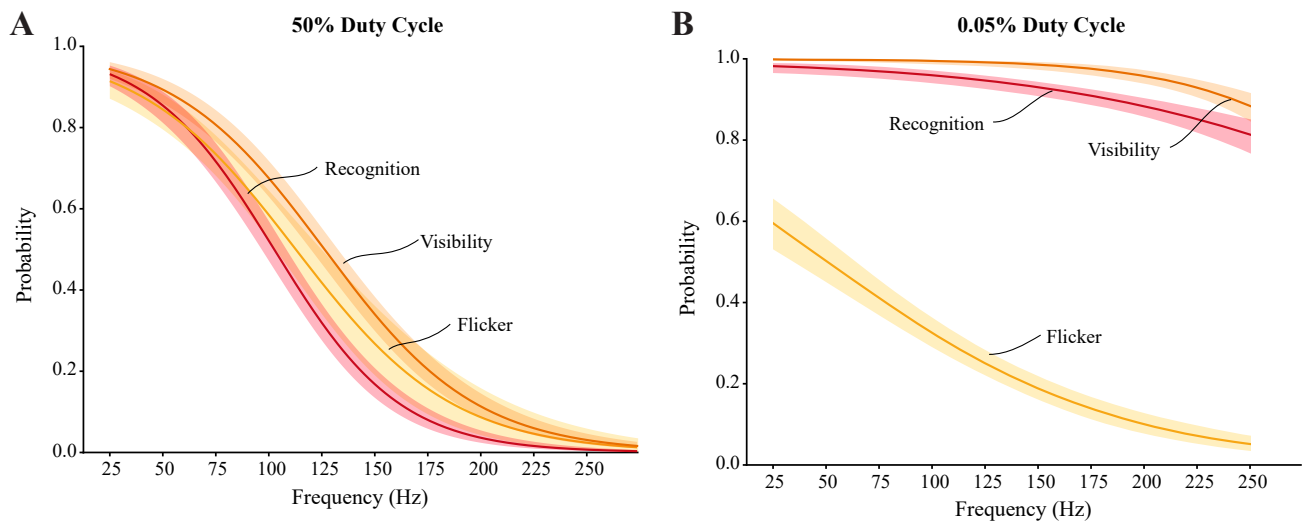

Figure 2. A. Luminance balanced letters were displayed with 50% duty cycle that provided bright and dark pulses of equal durations. Probability of perceiving flicker was high at 25 Hz, and this declined to zero for 250 Hz displays. Letters were only visible for displays where flicker could also be seen, and most letters that were visible could be identified. B. Displays that used 0.05% duty cycle, i.e., with ultrabrief bright pulses, had reduced levels of perceived flicker. Most letters were visible and were correctly identified even when there was no perceived flicker.

The strong correspondence of the three models (shown in Figure 2A) indicates that if a letter was visible, it could also be recognized. For these displays at 50% duty cycle, both states depended on presence of flicker. If the letters were not flickering, they merged with the background and became invisible.

Figure 2B provides the models for flicker perception, visibility, and letter recognition for the 0.05% duty-cycle displays. At 25 Hz, flicker was perceived on only 60% of the trials but the letters were visible (and recognized) for almost all displays. This means that about 40% of trials displayed at 25 Hz can be attributed to anomalous contrast. As was the case for 50% duty cycle, flicker perception declined monotonically as frequency was increased, reaching near zero at 250 Hz. By comparison, visibility and recognition of letter remained fairly high, declining by roughly 5-10% across the frequency range. Prior work found that recognition continues to decline at higher frequencies, but is still well above chance at 2000 Hz.[8] For most of the trials where flicker was not perceived, the letters were steady, visible, and could be correctly identified. The high visibility of non-flickering letters displayed at 0.05% duty cycle is a phenomenon that we are attributing to anomalous contrast. It is a special process that is specific to the luminance-balanced relationship, and was elicited by the use of ultra-brief bright pulses

For 50% and 0.05% models alike, probability of letter recognition was below visibility of the letters, progressively so as frequency was increased. This differential is due to the scoring of mis-identified letters, i.e., they would be marked as being visible but as not being correctly identified.

*3.4 Anomalous Contrast Effects as a Function of Display Cycles (Task 3)*

Thus far, this anomalous contrast has been demonstrated with 2 μs flicker that was displayed for 300 ms, this being 75 cycles at 4 ms per cycle. A number of successive flicker cycles may be needed to elicit the anomalous contrast effect, but we have no information on how many are required. To evaluate this issue, Task 3 displayed outline letters with bright-pulse durations of 2 μs or 2000 μs while varying the number of display cycles.

Figure 3A illustrates the light transitions for a given 2 μs and 2000 μs bright-pulse cycle. At 250 Hz, the duration of a bright-pulse cycle is 4000 μs. For the traditional 50% duty cycle with background being 8 Cd/m$^2$, the bright-pulse is an 8 Cd/m$^2$ incremental departure from background which lasts for half the cycle. The amount of energy the light adds is compensated exactly by the loss of light energy during the half-cycle in which no light is being emitted, i.e., the dark pulse. This counterbalance of light energy provides an average luminance that matches the background luminance. When the stimulus is displayed against a luminance-matched background, it is generally not visible.

For the 2 μs flicker, the duration of the dark pulse was 3998 μs, and the bright pulse provided a much higher intensity to maintain energy balance. A luminance-balanced letter that is displayed with this bright-pulse duration can elicit anomalous contrast if it is displayed for 300 ms (see Fig 1). The question here is how much letter recognition is elicited as a function of the number of bright-pulse cycles. Figure 3B shows the group models for judgments derived from Task 3. One can see that the 2000 μs means were at the chance level across the full range of cycles (only one subject was significantly above chance). However, recognition rose progressively for the 2 μs displays as the number of cycles was increased, reaching over 80% recognition by 32 cycles (128 ms). This may represent summation of weak activation that is being generated by each flicker cycle. Whatever the case, 32 cycles was deemed to be a convenient number to use for the tasks that followed.

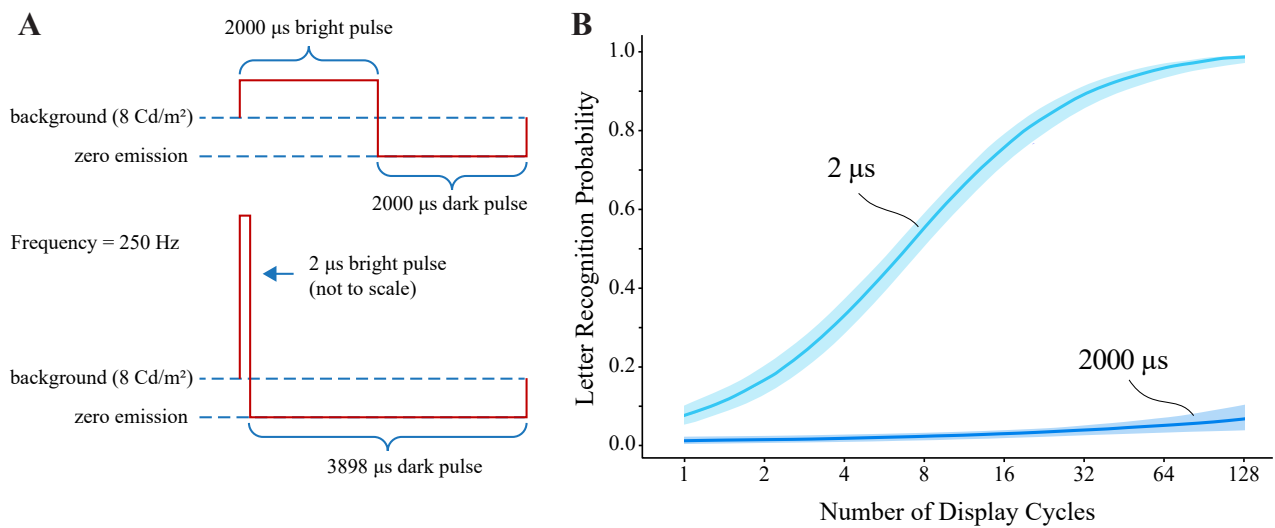

Figure 3. A. Standard flicker displays provide bright and dark pulses with equal amplitudes and duration. One can provide the same balance in light energy using a brief bright pulse that is more intense. B. Letters that were displayed with the standard bright/dark pulse ratio failed to rise much above chance performance. But the ultrabrief bright pulses produced a monotonic rise in recognition as the number of display cycles was increased.

### 3.5 Incremental Departure from Background (Task 4)

Task 4 adjusted bright and dark intensity levels of 250 Hz flicker, done with 2 μs flicker, with a given sequence being displayed for 32 cycles. The size of bright and dark pulses was varied, these being registered as departures from background intensity. The reductions were scaled with respect to range available to dark pulses, with 0% being no departure from background intensity and 100% being the absence of light emission. Bright-pulse intensity was correspondingly adjusted to maintain luminance balance relative to background.

Figure 4A illustrates bright and dark amplitudes for a given flicker cycle, and the model of judgments can be seen in Figure 4B. Anomalous contrast recognition was minimal at 10%, but it climbed quickly and reached an asymptote at about 40% departure from background intensity. These result quantify the role of net departure from background intensity level that can elicit perception of anomalous luminance contrast, and indicate that activation of anomalous contrast asymptotes early, i.e., being especially sensitive to the small departures from background, with larger departures contributing less to the activation. This kind of non-linearity is not uncommon for physiological activation that is based on sharp thresholds within a pool of mutually-responding elements. As one passes the threshold levels, there is an avalanche of activation within the pool.

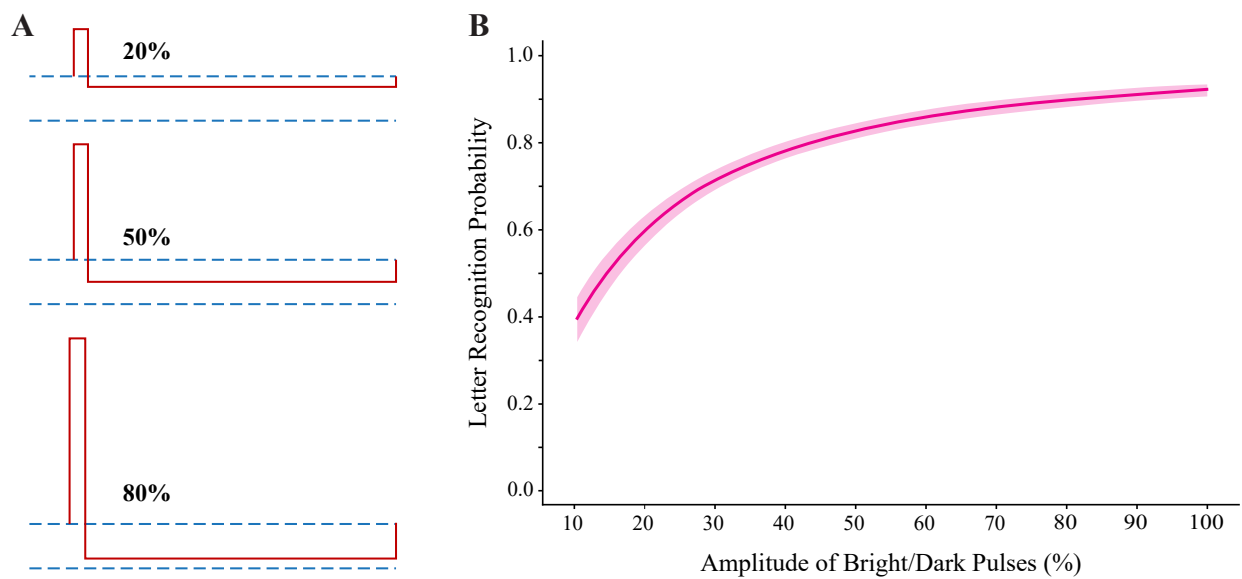

Figure 4. A. One can evaluate how the probability of flicker-fused letter recognition changes as amplitude of departure from background is increased. The amplitude can be scaled as a percent change in the dark pulse, with amplitude of the bright pulse being adjusted to maintain an average luminance that matches the background. B. Probability of recognition rose as the size of departure from background increased, with the fastest rise being at the lower end of the intensity departure scale.

### 3.6 Abbreviated Dark Component -- Bright First (Task 5)

This protocol also adjusted bright and dark intensity levels of 250 Hz flicker, done with 2 μs flicker, with a given sequence being displayed for 32 cycles. The task was designed to evaluate the role of duration of the dark pulse. Dark-pulse intensity was 100% (no light emission), and duration was specified as a percentage of flicker period. This provides equivalent percentage scaling of bright- and dark-pulse activation for Tasks 4 and 5, as well as Task 6 that follows). The dark-pulse came immediately after the bright pulse but returned to background intensity at the end of the specified duration. The protocol provided 10 levels of dark-pulse duration, from 10% to 100%. Transitions within a flicker cycle are illustrated in Figure 5A.

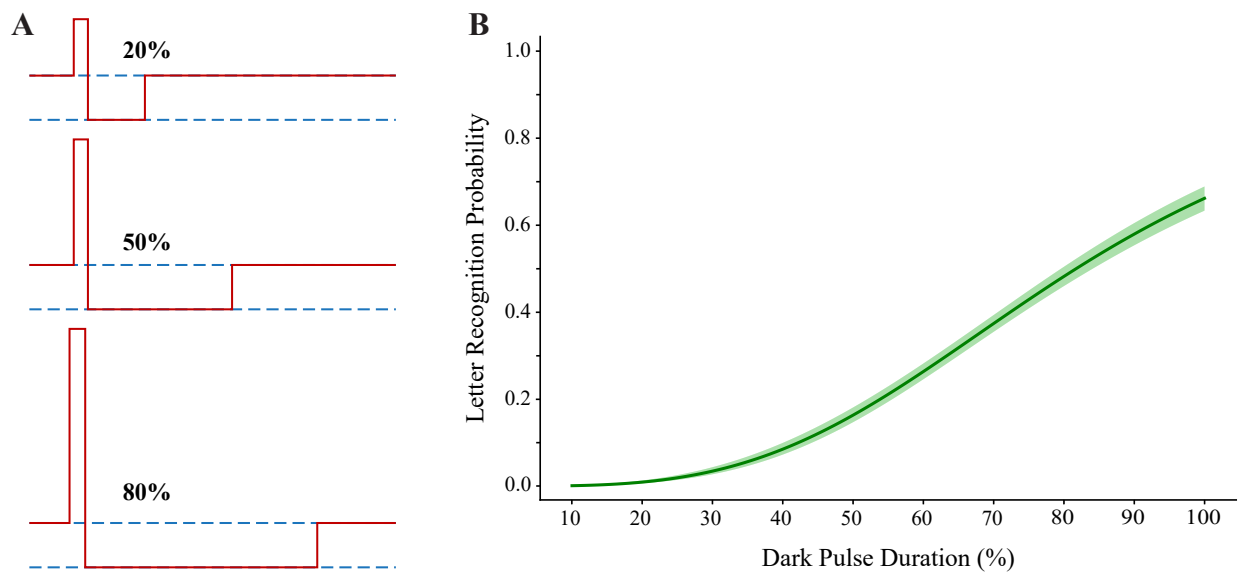

Figure 5. A. One can provide 100% dark-pulse departures from background as a percentage of flicker cycle, providing bright- and dark-pulse activations matching the scale used for Task 3. B. The model for probability of recognition rose slowly with increments of dark duration, which is unlike the profile of the model shown in Figure 4B.

Figure 5B shows a logistic regression model for group means. Unlike what was seen with Task 4, the initial increments of dark duration produced a slow rise in probability of recognition. Whereas Task 4 was yielding about 80% recognition when the departure from background was at 50%, the hit rate was still only about 25% with a dark duration that provided the same net energy.

### 3.7 Abbreviated Dark Component -- Dark First (Task 6)

To more fully test how positioning of the dark pulse might affect recognition, Task 6 reversed the order, presenting the pulse before the bright pulse rather than after. This is illustrated in Figure 6A. The protocol adjusted bright and dark intensity levels of 250 Hz flicker, done with 2 μs bright-pulses. It used the same ten levels of dark-pulse duration that were provided in Task 5, each trial displaying 32 cycles of flicker.

Respondents that were tested with Task 6 provided the regression model shown in Figure 6B. The probability of recognition rose rapidly as the duration of the dark pulse was increased from 10% to 50%, with the hit rate being above 80% for the 50% durations. This is similar to what Task 4 provided (Fig 4B). The rapid rise that was manifested when the dark pulse preceded the bright pulse differs greatly from the late-rising profile that was found where the dark pulse followed the bright pulse (Fig 5B). Clearly the order of bright- and dark-pulse display is a critical factor in whether a luminance balanced flicker-fused letter is visible, and therefore available for recognition. This unexpected order effect will be discussed below.

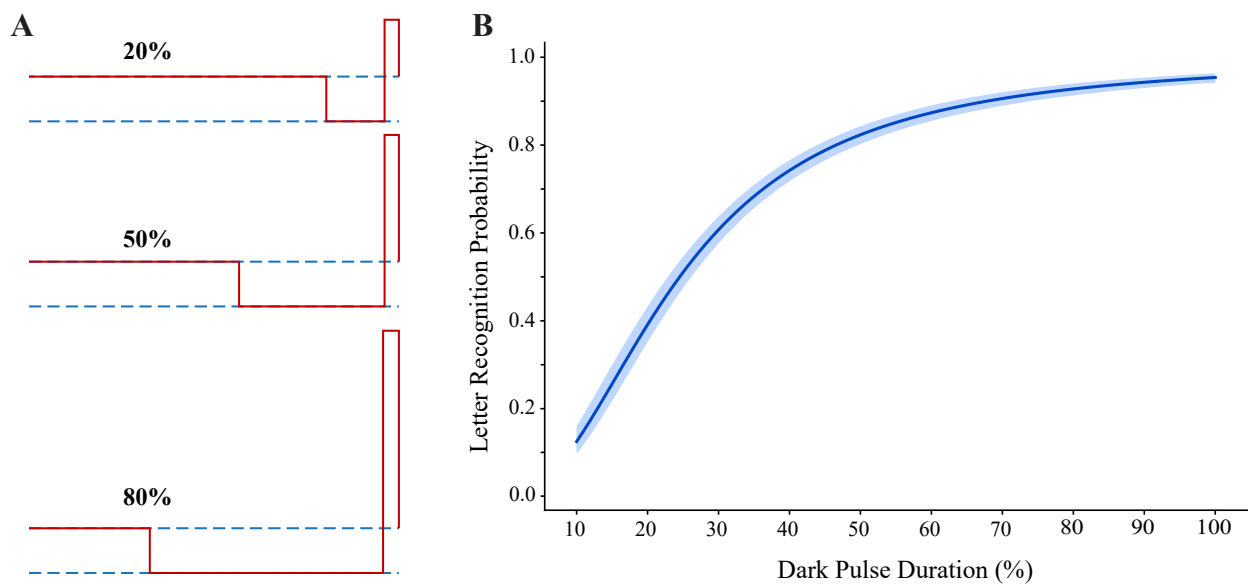

Figure 6. A. For each flicker cycle, the increments of dark duration were displayed in advance of the 2 μs bright pulse. B. Probability of letter recognition rose rapidly, providing a hit rate in the 80% range with a dark duration at 40% of the full flicker period. This is like the model profile that can be seen in Figure 4B, and is unlike what can be seen in Figure 5B.

### 3.8 Eliciting Anomalous Contrast Recognition with Movement (Task 7)

Thus far, we have used ultra-brief bright pulses to elicit anomalous contrast, which provides for visibility of luminance-balanced letters. The alignment of dots in a given letter pattern determines whether it can be identified, and we have been assuming that the ultra-brief pulses activate retinal circuits that register those alignments. Numerous theorists have suggested that eye and object motion is critical for registering alignment of contours, and one might logically include alignment of dots. So we thought it would be useful to examine whether motion of the letter pattern could substitute for ultra-brief bright-pulses and provide visibility (and recognition) of luminance-balanced letters.

Following the basic Task 1 protocol that provided the models shown in Figure 1, we displayed flicker-fused letters across a range of contrasts that bracketed the luminance-balance level (100%).  But unlike Task 1, which displayed half of the letters with 2 µs flicker, here all the letters were displayed with a 2000 µs flicker on every trial of the task.  Motion substituted for the 2 µs condition, wherein the letter was shifted back and forth to produce what can be described as "jiggle".  The "still" control condition did not change the position of the letter across the 300 ms display interval.  One group of respondents judged solid-fill Arial 60 letters, and a second group judged the outline Arial 60 letters.  The logistic group models for probability of letter recognition as a function of contrast for these treatment combinations are provided in Figure 7A and 7B.

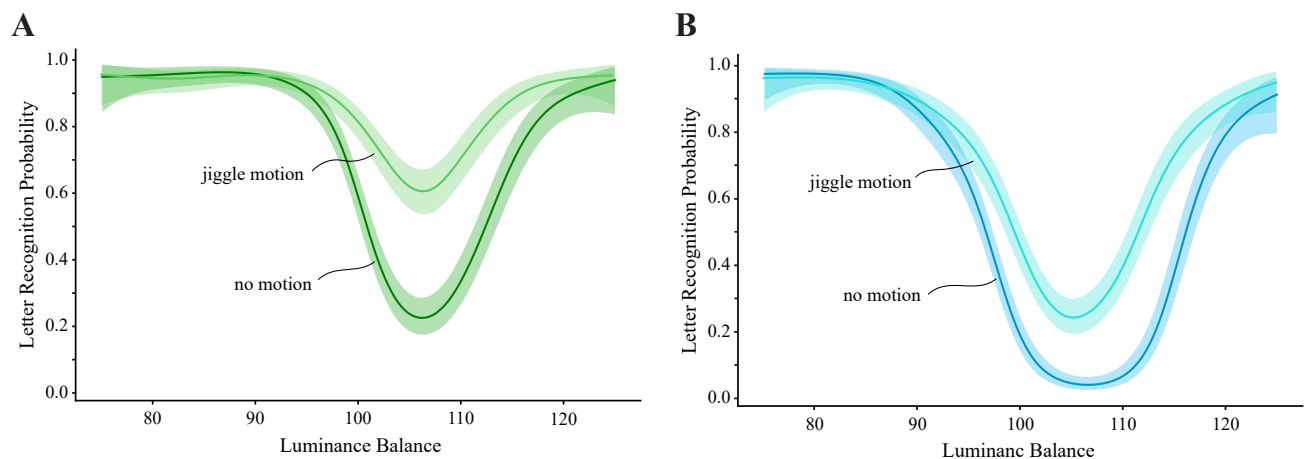

Figure 7.  A. Solid-fill letters that were displayed without any motion manifested the expected reduction of recognition in the vicinity of luminance balance.  Jiggling the letters greatly increased recognition, i.e., the motion produced anomalous contrast.  B. Motion vs. non-motion produced similar effects for outline letters.  The anomalous contrast effect was weaker, possibly because the merge with background was stronger, as manifested in the absence of motion.

## 4. DISCUSSION

Our focus is on retinal mechanisms that register and signal luminance and luminance contrast, and more specifically those that can discriminate contours that match background luminance.  We expect that a letter-shaped dot pattern that differs in its luminance relative to background will be visible, and potentially recognizable.  It is strange that luminance-balanced letters that should be invisible can be identified when ultrabrief bright pulses are used to produce the flicker.[7,8]  This anomalous contrast effect was found here again as a difference in probability of recognition for letters displayed with 2 µs flicker versus 2000 µs flicker (see Figs 1 and 3).  Most letters that were displayed with 2 µs flicker could be identified when their average luminance matched background luminance, a condition wherein one would expect them to be invisible.

To explain the basis for their visibility, we adopt a "dual-channel hypothesis" with respect to retinal mechanisms. This hypothesis, as well as the most relevant retinal physiology, is discussed elsewhere.[12] What follows is a general outline of the concepts.

We believe that the present findings reflect activity of retinal circuits that encode and signal luminance contrast of figures, relative to background luminance. Key components for these mechanisms include the ON, OFF, and ON/OFF bipolar cells that register activation of photoreceptors, and ganglion cells that signal that activity to other parts of the brain.[6] Horizontal and amacrine cells contribute to the encoding, but for present purposes it isn't necessary to spell out their role.

The first step is the encoding of luminance level of a stimulus, wherein the degree of photoreceptor activation is passed to ON and OFF bipolar cells. When a bright light level is driving high levels of tonic activity in the ON bipolar cells, it is producing a low level of tonic activity in OFF bipolars. This activity is conveyed to ON and OFF ganglion cells (and their optic nerve fibers), with relatively high and low firing rates reflecting the level of stimulus luminance.

The dual-channel hypothesis asserts that a steady stimulus will be registered by both channels. Note that because of a signal inversion that takes place at the photoreceptor/bipolar synapse, a dark stimulus produces a high firing rate in the dark channel and a low firing rate in the bright channel. Activity levels are reciprocal, and the specific level of brightness that is consciously perceived makes use of both signals.

Luminance of a uniform zone is signaled by the combined activity of dark and bright channels. Luminance contrast is manifested when the light activation for a zone differs from the activity level of the surrounding area. "Receptive field" mechanisms encode and signal these differentials in energy balance. A given bright or dark channel has a central zone that is registered the light level, but does so in relation to a larger surrounding zone. This is essentially a figure/ground matter, with retinal circuits registering the luminance contrast of the figure relative to background luminance.

A bit of complexity is added to the story when a flicker sequence provides the figure, e.g., a letter pattern. Here the stimulus alternates between being bright and dark, but it appears steady (fused) if the frequency is high enough that the individual pulses cannot be registered. The perceived brightness of the zone will depend on the relative light energy being provided by the bright and dark pulses. The Talbot-Plateau law specifies what combination of pulse intensities, frequency, and duration of a flicker-fused figure will match the steady brightness (luminance) of the background.[7,8] If one tips the balance to have more energy in the bright-pulse portion of the flicker cycle, one will see a bright figure against the background. Tipping the energy balance to increase dark-pulse energy yields a dark figure. The system is very sensitive to the balance, and a small differential can affect probability of letter recognition.[8]

Figure 2A shows that for a 50% duty cycle, i.e., equal duration of bright and dark pulses, luminance-balanced letters disappear as soon as one can no longer register the individual pulses of the flicker. The recognition model had slightly lower probabilities, the difference most likely being from mis-identification of letters that added 0s to the model. But for the 0.05% duty cycle the vast majority of the letters were visible and were identified rather than merging with background to become invisible (see Fig 2B). This provides a dramatic example of the anomalous contrast effect.

We submit that using pulse-pairs that have high equal-but-opposite energy (intensity x duration) will block merging of dark- and bright-channel activity that would normally produce invisibility. This is the case for the 0.05% duty cycle, where the bright pulse if brief but very bright, and the dark pulse is at maximum departure and has a duration that is almost the full cycle period. So even though the letter is physically luminance balanced (relative to background), a dark or bright channel becomes dominant, and the stimulus, e.g., letter, is made visible.

We return to the question of what retinal mechanisms register a figure as being distinct from the background. The edge of a figure might activate an array of receptive-field centers, signaling them as being distinct from the background. Further, the mechanism might be especially sensitive to the timing of center activation in relation to the surround. Motion of the figure may be especially effective in eliciting a combined response from the receptive-field array. The models in Figures 5 and 6 show that that a dark pulse is more effective when it precedes the bright pulse rather than following it. A moving contour would encounter the surround before reaching the center of the receptive field, so an anomalous contrast trigger might be most sensitive to this successive activation.

The intuition that motion circuits might be involved in producing anomalous contrast provided the basis for conducting Task 7. Additional experiments would be needed to more fully document the motion effects, but the initial finding suggests that ultrabrief bright pulses are activating motion sensors. Direction-selective ON/OFF ganglion cells may be especially relevant. Hanson and associates [13] have reported that these cells are also orientation selective -- they respond more vigorously to an edge that moves into their receptive field. The same laboratory subsequently extended the initial findings, providing evidence of selective response to horizontal as well as vertical orientations.[14] The evidence suggests that "starburst" amacrine cells are providing the selective control of ganglion cell activity. It is especially relevant that the orientation-specific responses can be elicited by static displays. In other words, even though direction-selective ON/OFF ganglion cells are generally considered to be motion sensors, they could register the orientation of fixed-location contours that were displayed with a brief pulse or pulse sequence.

The fact that ON/OFF ganglion cells register both increments and decrements of light would make them especially useful for encoding contours irrespective of their contrast polarity. If the receptive-field center can be activated by a bright pulse or a dark pulse, the fact that our 2 μs pulse was bright becomes moot. Further, polarity of the surround would also be moot, but duration of the pulse elements might be relevant. Contours that passed first into the larger surround would be delivering long-duration influence, followed by a brief spike in activation as the smaller center was encountered. This might explain the differential order effects that were found in Tasks 5 and 6.

5. CODA

The Talbot-Plateau law specifies the brightness that will be perceived when a flickering stimulus is produced by a given combination of pulse intensities, durations, and frequency. When a combination provides stimulus brightness that matches brightness of the background within which it is embedded, the stimulus should be invisible. We view this as a natural extension of the Talbot-Plateau law, and if the combination produces a visible stimulus, it is violating the law. This might be an adaptive override of the retinal mechanisms that combine and signal the luminance average.

Natural images are seldom uniform in luminance, but are filled with irregularities that can be collectively described as texture. The ability to perceive a figure can be based on how its texture differs from its background. But even where the figure and background have the same texture, coordinated movement of the texture elements can make the figure appear. Investigators have long been aware of how camouflage can be disrupted by motion.[15-18]

What may be less obvious is how the receptive fields of retinal channels would be affected by movement of the figure texture. Depending on speed of motion and density of the texture, there will be fluctuation of stimulation as dark and light portions of the texture passed across the receptive field. Those fluctuations could also be described as flicker, and to complete the circle, we are suggesting that some flicker combinations can register and signal the presence of figural boundaries due to the coordinated movement of texture elements.

The form-from-motion that can be produced when figure and background have matching textures provides a convenient illustration of how a violation of the Talbot-Plateau law can provide adaptive benefit. But such a mechanism would have more general utility in boosting discrimination of form boundaries even where the textures differ. The visual system makes use of a number of different tools for accomplishing figure/ground discrimination. Adding an anomalous contour mechanism that registers and signals fluctuations of receptive-field stimulation would be a useful addition to the toolbox, even if it requires blocking of signals that are designed to report the average luminance.

**Acknowledgments**     Jack Shields and Viktor Pham conducted testing of respondents and curation of data. Regression modeling was provided by Wei Wang, and Manhui Zhu. Figure graphics were constructed by Lynn Tu.

**Funding**   Support for this research was provided by the Neuropsychology Foundation and the Quest for Truth Foundation.

**Data Accessibility Statement.**   The data that support the findings of this study are openly available at:

https://urldefense.com/v3/__https://doi.org/10.6084/m9.figshare.29262089__;!!LIr3w8kk_Xxm!qDOmI5aX2iyuKg6M38MJS3tnv2gP2xqjLHnmroxTmm9HgYo8nWNj546iYOa5cWNwHyc9DgvbXA$

DOI reference -- 10.6084/m9.figshare.29262089

**Author Contributions.**  Both authors had full access to all the data in the study and take responsibility for the integrity of the data and the accuracy of the data analysis.
*Conceptualization*:  EG and JM;  *Methodology*: EG and JM;  *Equipment design, fabrication, and software*: JM;  *Funding and Supervision*: EG;  *Writing - Original draft*:  EG;  *Writing- Review & Editing*:  EG and JM.